\documentclass{bvm2022} % Do NOT change this line
\makeatletter
\let\blx@rerun@biber\relax
\makeatother
\usepackage[misc]{ifsym}

\begin{document}
\definecolor{myYellow}{rgb}{0.93,0.84,0.57}
\definecolor{myBrown}{rgb}{0.71,0.54,0.46}
\definecolor{myBlue}{rgb}{0.59,0.72,0.84}
% Do not change anything in the preamble (anything above \begin{document}) except for the specification of the bibliography file, any additional changes will be lost

% use the \selectlanguage command to select the language in which your proceedings are written

%\selectlanguage{ngerman} % German
\selectlanguage{english} % English

% Indication of the title of your contribution
\title{Few-shot Unsupervised Domain Adaptation for Multi-modal Cardiac Image Segmentation}
% If you write a short paper/abstract, the title must start with "Abstract:".
% \title{Abstract: Bildverarbeitung für die Medizin 2022}

% Optional specification of subtitle
% \subtitle{Guidelines for the creation of the print-ready contributions}

% titlerunning appears in the header of every second page
% LaTeX generates this automatically from your contribution title
% However, if it is too long, the message "Title Suppressed Due to Excessive Length" appears instead.
% In this case, specify an abbreviated form of the title here
\titlerunning{FUDA for Multi-modal Cardiac Image Segmentation}

% Please indicate all authors involved
% To allow us to correctly identify the last name of each author, indicate it using the \lname{} command.
% If more than one institute is involved, list the number of the institute(s) (see below) with \inst{} after the respective author. If only one institute is involved, omit this.
% Separate all authors with a comma
\author{Mingxuan \lname{Gu}$^1$(\Letter), Sulaiman \lname{Vesal}$^1$, Ronak \lname{Kosti}$^1$, Andreas \lname{Maier}$^1$}

% Enter the authors here as you want them to appear in the header
% Name only the surnames
% Depending on the number of authors involved, follow the examples below
% \authorrunning{Meier} - one author
% \authorrunning{Meier \& Müller} - two authors
% \authorrunning{Meier, Müller \& Schulze} - three authors
% \authorrunning{Meier et al.} - more than three authors
\authorrunning{Gu et al.}

% Specify the institutes involved
% In case of participation of more than one institute, each institute shall be preceded by an ascending number with \inst{}.
% If only one institute is involved, omit the corresponding number.
% Separate individual institutes with \\
\institute{
$^1$Pattern Recognition Lab, Friedrich-Alexander-Universit\"at Erlangen-N\"urnberg (FAU), Erlangen, Germany}

% Enter the e-mail address of the corresponding author
\email{mingxuan.gu@fau.de}

\maketitle

% Abstract of your contribution, only for long contributions
% Do NOT use \begin{abstract} ... \end{abstract} for short articles
\begin{abstract}
Unsupervised domain adaptation (UDA) methods intend to reduce the gap between source and target domains by using unlabeled target domain and labeled source domain data, however, in the medical domain, target domain data may not always be easily available, and acquiring new samples is generally time-consuming. This restricts the development of UDA methods for new domains. In this paper, we explore the potential of UDA in a more challenging while realistic scenario where only one unlabeled target patient sample is available. We call it Few-shot Unsupervised Domain adaptation (FUDA). We first generate target-style images from source images and explore diverse target styles from a single target patient with Random Adaptive Instance Normalization (RAIN). Then, a segmentation network is trained in a supervised manner with the generated target images. Our experiments demonstrate that FUDA improves the segmentation performance by 0.33 of Dice score on the target domain compared with the baseline, and it also gives 0.28 of Dice score improvement in a more rigorous one-shot setting. Our code is available at \url{https://github.com/MingxuanGu/Few-shot-UDA}.
\end{abstract}

\section{Introduction}

As manual contouring of medical images is tedious and time-consuming, automatic medical image segmentation is more desirable~\cite{3260-01}.
While deep learning methods often suffer from performance degradation when a domain gap is observed between training (source) and testing (target) data. UDA methods tackle this problem by reducing the domain gap with a variety of techniques, for example, discrepancy reduction~\cite{3260-02}, adversarial learning~\cite{3260-03}, image translation~\cite{3260-04}, etc. These methods are conditioned on the availability of a large amount of target data which, however, is quite scarce. 

In this work, we consider a more realistic and practical scenario where we still have sufficient labeled source data, while we only have one unlabeled target data for training. To this end, a style transfer method called Random Adaptive Instance Normalization (RAIN)~\cite{3260-05} is used to generate diverse target-style images from a single target patient data. Then, a segmentation module can be trained in a supervised manner with generated images. Our contributions are: (1) we explore the potential of FUDA for multi-modal cardiac CMR segmentation and it shows better performance compared with its baseline model and other recent UDA methods, (2) we extend our method to one-shot learning and demonstrate the possibility of FUDA with only one slice of target data available.

\begin{figure}[bt]
\includegraphics[width=\textwidth]{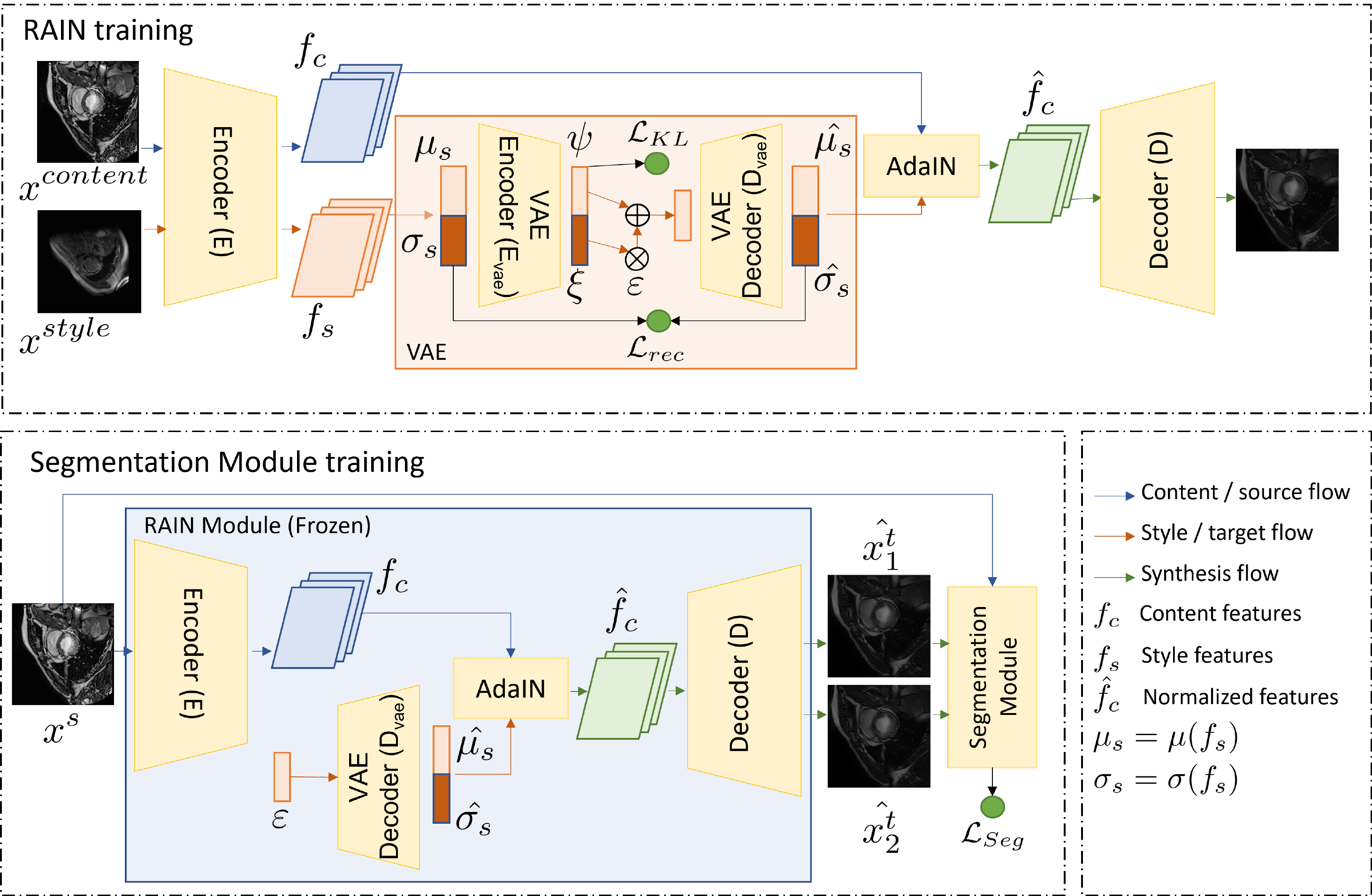}
\caption{Overview of the proposed FUDA segmentation framework. RAIN is first pre-trained with $x^{content}$ (bSSFP) and $x^{style}$ (T2). Then during training of the segmentation module, stylized target images will be generated by RAIN with source images and $\varepsilon$ generated from target image(s). After that, the segmentation module can be trained using the source and stylized images. The $\varepsilon$ is iteratively updated to generate more difficult stylized images.} 
\label{3260-network_arc}
\end{figure}

\section{Materials and methods}
\subsection{Dataset}
We assess the proposed FUDA on MS-CMRSeg~\cite{3260-06} 2019 challenge dataset, which consists of 45 short-axis bSSFP, T2-weighted and LGE scans from patients diagnosed with cardiomyopathy. The ground-truth contours are generated by two experts and include right ventricle (RV) cavity, left ventricle (LV) cavity, and myocardium (Myo) region. Only affine transformation (rotation, translation, shearing, etc.) is applied and the sequences are normalized using min-max normalization. Then, the sequences are center-cropped to $224\times224$ pixels to have only region-of-interest (ROIs) areas.

\subsection{Problem statement} In UDA for semantic segmentation, a set of labeled data in source domain $\mathcal{D}^{s}(x^{s}, y^{s})$ is given, where $x^{s}$ represents one sample, and $y^{s}$ the corresponding label in $\mathcal{D}^s$. Whereas for target domain ($\mathcal{D}^{t}$) only unlabeled target images ($x^{t}$) are given. The goal is to improve the performance of the segmentation by reducing the distribution gap between the source and target domain. In our case, we consider only one unlabeled target patient data (${x}^{t}$) being available.

\subsection{RAIN Module}
RAIN is developed on the basis of Adaptive Instance Normalization (AdaIN)~\cite{3260-07}. AdaIN has an encoder-decoder architecture. It generates stylized images which have the appearance of the style image while preserving the structure of the content images by re-normalizing the features of content images with Eq. ~\ref{3260-adain}:

\begin{equation}
    AdaIN(f_{c},f_s) = \sigma(f_s) (\frac{f_{c}-\mu(f_{c})}{\sigma(f_{c})}) + \mu(f_s),
    \label{3260-adain}
\end{equation}
where $f_c$, $f_s$ are the latent features of the content and style image, $\mu(.)$ and $\sigma(.)$ denote channel-wise mean and standard deviation. To achieve realistic image transfer, a content loss $\mathcal{L}_c$ and a style loss $\mathcal{L}_s$~\cite{3260-07} is employed.

RAIN takes advantage of the style transfer in AdaIN and involves a style variational auto-encoder (VAE) in between the encoder and the decoder. The style-VAE is composed of an encoder $E_{vae}$ and a decoder $D_{vae}$. 
$E_{vae}$ encodes $\mu(f_{s}) \oplus  \sigma(f_{s})$ to $N(\psi, \xi)$, where $\oplus$ denotes concatenation, and $D_{vae}$ aims to reconstruct the original style with $\widehat{\mu(f_s) \oplus \sigma(f_s)}=D_{vae}(\varepsilon)$, where $\varepsilon\sim N(\psi, \xi)$. Kullback-Leibler (KL) divergence loss is applied to enforce $N(\psi, \xi)$ to be normal distributed. Furthermore, L2 loss is applied between $\mu(f_{s}) \oplus  \sigma(f_{s})$ and $\widehat{\mu(f_s) \oplus \sigma(f_s)}$ to force identity reconstruction $\mathcal{L}_{Rec}$.
 
Thus, the overall loss function to train RAIN can be formulated as $\mathcal{L}_{RAIN} = \mathcal{L}_{c} + \lambda_{s}\mathcal{L}_{s} + \lambda_{kL}\mathcal{L}_{KL} + \lambda_{Rec} \mathcal{L}_{Rec}$
% \begin{equation}
%     \mathcal{L}_{RAIN} = \mathcal{L}_{c} + \lambda_{s}\mathcal{L}_{s} + \lambda_{kL}\mathcal{L}_{KL} + \lambda_{Rec} \mathcal{L}_{Rec}
% \end{equation}
where $\lambda_s$, $\lambda_{KL}$ and $\lambda_{Rec}$ denote the weights for the losses.
\begin{figure}[bt]
\includegraphics[width=\textwidth]{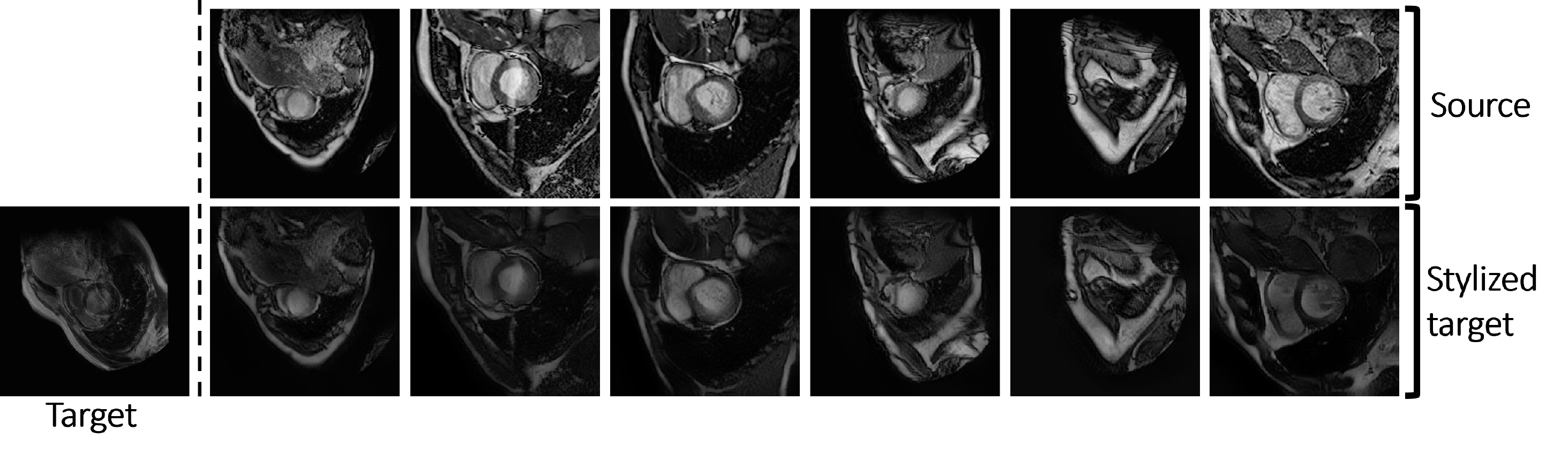}
\caption{Qualitative results of RAIN on bSSFP $\rightarrow$ LGE. The first column shows a LGE axial slice as the style image. The first row shows the bSSFP images as the content images, the second row shows the corresponding stylized LGE images.} 
\label{3260-rain_style}
\end{figure}
\begin{figure}[t]
    \centering
    \includegraphics[width=\textwidth]{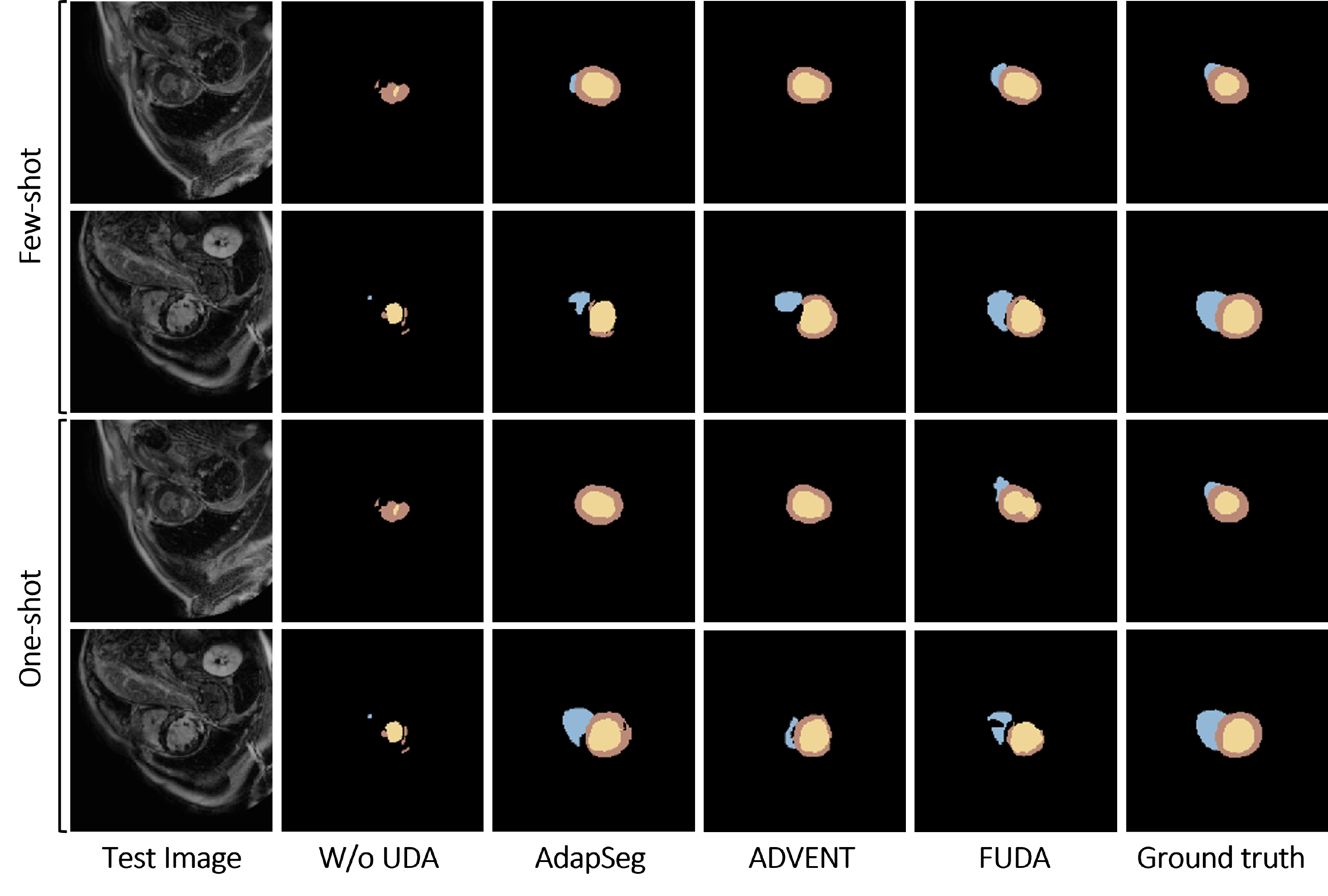}
    \caption{A visual comparison of segmentation output results produced by different methods for test LGE images under one and few-shot learning settings. The first two rows show the results for few-shot models and the last two rows are for one-shot respectively. LV is shown in \textcolor{myYellow}{yellow}, Myo in \textcolor{myBrown}{brown} and RV in \textcolor{myBlue}{light-blue} colors.}
    \label{3260-fewshot}
\end{figure}
\subsection{Few-shot UDA for cardiac MRI segmentation}
The proposed FUDA framework is constructed based on ASM~\cite{3260-05} implementation and is shown in Fig.~\ref{3260-network_arc}. To train the segmentation module and explore styles from a single target image, we first generate the latent distribution $\mathcal{N}(\psi, \xi)$ from the input image $x^t$. Then we sample an $\varepsilon_0$ from $\mathcal{N}(\psi, \xi)$. After that, we can generate stylized images $\hat{x^t}$ from any source image $x^s$. 
% For convenience, we rewrite the AdaIN formula as:
% \begin{equation}
%     AdaIN(f_c, \mu_s\oplus\sigma_s) = \sigma_s \frac{f_c - \mu(f_c)}{\sigma(f_c)} + \mu_s
% \end{equation}
Then the style transfer can be formulated as:
% $\hat{x^t} = D(AdaIN(E(x^s), D_{vae}(\varepsilon)))$.
\begin{equation}
    \hat{x^t} = D(AdaIN(E(x^s), D_{vae}(\varepsilon)))
\end{equation}
Furthermore, the segmentation module can be trained in a supervised manner with $x^s$ and the corresponding $\hat{x^t}$. 

We train the segmentation module with a combination of cross-entropy loss (CE) and jaccard distance loss (JD) as $\mathcal{L}_{seg} = \mathcal{L}_{CE} + \mathcal{L}_{JD}$.
To enforce the segmentation module to produce domain invariant features between $x^s$ and the corresponding $\hat{x^t}$, $\mathcal{L}_{con} = ||z^s - z^t||_{2}$ is applied as a consistency loss,
where $z^{s/t}$ are the latent features of the source and the corresponding generated target image from the bottleneck layer of the segmentation module. Then the overall loss function for the segmentation module is:
\begin{equation}
    \mathcal{L}_S = \mathcal{L}_{seg} + \lambda\mathcal{L}_{con}
\end{equation}
where $\lambda$ is the weight of $
\mathcal{L}_{con}$. 
Finally, to generate more diverse and increasingly difficult target images , $\varepsilon$ is updated in the direction that makes the segmentation module perform worse on segmentation with:
\begin{equation}
    \varepsilon_{i+1} = \varepsilon_i + \alpha\triangledown_{\varepsilon_i}\mathcal{L}_{seg}(\hat{x^t},y^s)
\end{equation}
where $i$ is the iteration number, $y^s$ represents the corresponding source label for $\hat{x^t}$ and $\alpha$ denotes the learning rate.
\begin{table}[t]
\caption{Dice coefficient (DC) and Hausdorff distance (HD) measures for the proposed FUDA together with baseline (W/o UDA) method, inter-observer study and the performance of the two successful UDA methods AdaptSeg~\cite{3260-08} and ADVENT~\cite{3260-03}. Baseline is trained with bSSFP images and tested directly on LGE images. For few-shot learning, we take the whole LGE squence of patient 10 as the training target data. For one-shot learning, we only use slice 13 of patient 10 as the training target data. Best results are shown in bold.}
\label{3260-udaresults}
\begin{tabular*}{\textwidth}{l@{\extracolsep\fill}lllll|lllll}
\hline
         &      & \multicolumn{4}{c}{\textbf{DC ($\uparrow$)}}  & \multicolumn{4}{c}{\textbf{HD\ts[mm] ($\downarrow$)}}     \\ \hline
\textbf{Method }  & \textbf{Data} & \textbf{Myo}  & \textbf{LV}   & \textbf{RV}   & \textbf{AVG } & \textbf{Myo}  & \textbf{LV }  & \textbf{RV}    & \textbf{AVG } \\
\hline
W/o UDA  & N/A  & 0.24 & 0.40 & 0.27 & 0.30 & 31.7 & 31.0 & 45.0  & 35.9 \\
\hline
AdaptSeg & Few  & 0.39 & 0.63 & 0.58 & 0.53 & 39.1 & 28.7 & 25.6  & 31.1 \\
ADVENT   & Few  & 0.39 & 0.59 & 0.52 & 0.50 & 38.4 & 35.3 & 37.9  & 37.2 \\
Proposd  & Few  & \textbf{\emph{0.46}} & \textbf{\emph{0.77}} & \textbf{\emph{0.65}} & \textbf{\emph{0.63}} & \textbf{\emph{24.5}} & \textbf{\emph{13.7}} & \textbf{\emph{22.4}}  & \textbf{\emph{20.2}} \\
\hline
AdaptSeg & One  & \textbf{\emph{0.45}} & 0.65 & 0.52 & 0.54 & \textbf{\emph{32.3}} & 34.3 & 36.4  & 34.3 \\
ADVENT   & One  & 0.37 & 0.61 & 0.51 & 0.50 & 42.7 & 26.9 & 35.0  & 34.9 \\
Proposd  & One  & 0.39 & \textbf{\emph{0.73}} & \textbf{\emph{0.63}} & \textbf{\emph{0.58}} & 36.2 & \textbf{\emph{19.0}} & \textbf{\emph{25.6}}  & \textbf{\emph{26.9}} \\
\hline
Observer & N/A  & 0.76 & 0.88 & 0.81 & 0.82 & 12.0 & 14.3 & 21.5  & 15.9 \\ \hline
\end{tabular*}
\end{table}

\subsection{Training} The training process has two stages. First, we use bSSFP-MRI as content images and T2-weighted-MRI as style images to pretrain the RAIN module. Since it does not involve any target images (LGE), this process makes it convenient to train RAIN anytime before training the segmentation module. In the second stage, the RAIN model is frozen, and we employ Dilated-Residual UNet (DR-UNet)~\cite{3260-09} as the segmentation module. We first pretrain the DR-UNet with bSSFP images and the generated LGE-style images in a supervised manner for 40-50$k$ iterations. Then the model is trained together with $\varepsilon$ being iteratively updated for another 40-50$k$ iterations. We used stochastic gradient descent (SGD) with a momentum of $0.9$ and a weight decay of $5e-4$ as the optimizer. We empirically set the hyper-parameters of $\lambda_s = 5, \lambda_{KL} = 1$, $\lambda_{Rec} = 5$ and $\lambda = 2e-3$. The proposed method is trained and tested on one Geforce 1080Ti GPU. The training of RAIN and segmentation module takes 2 hours and 21 hours respectively. Overall inference takes 23 seconds in average for each patient.

\section{Results}
Fig.~\ref{3260-rain_style} shows the qualitative results of the pretrained RAIN. We can observe that RAIN successfully captures the features of target images. Tab.~\ref{3260-udaresults} summarizes the quantitative results of different methods. Baseline method achieved the lowest average volumetric Dice (0.30) and an HD (35.9\ts mm). With a few-shot UDA setting, our proposed  method achieved the best overall Dice (0.63) and the best lowest HD (20.2\ts mm). Subsequently, we demonstrate the results for one-shot UDA, and our proposed method achieved the highest Dice (0.58) and the lowest HD (26.9\ts mm). Fig.~\ref{3260-fewshot} illustrates qualitative examples of different segmentation approaches. Compared with other methods, the proposed method is able to produce more complete and precise segmentation maps.

\section{Discussion}
In this work, we presented a few-shot UDA (FUDA) for multi-modal CMR image segmentation while restricting the experiments in a more challenging yet realistic scenario where only one target sample is available. By comparing the proposed method with other approaches under the same settings, we show that FUDA highly reduced the domain gap with only a few target slices. We also demonstrated that the proposed few-shot method produces promising results with the more rigorous one-shot setting. We find for conventional UDA methods like AdaptSeg and ADVENT, performance on one-shot and few-shot settings only has a slight difference. This can be attributed to the fact that the slices of one patient only have a small distribution shift, hence the knowledge learned by the model from target slices of one patient is limited. While for FUDA, the model has the ability to explore unseen styles of the target images, hence more data provided results in more diverse target styles. Consequently, better segmentation performance could be achieved. Furthermore, we believe there is still room to improve the quality of the segmentation prediction. As a result, in the future, we will explore feasible techniques like contrastive learning and attention to improve the performance of the proposed FUDA.

% This command generates the bibliography using the entries of the .bib file.
% Remove it only if you do not use a bibliography. 
\printbibliography

@article{3260-01,
  title={Fully automatic segmentation of left ventricular anatomy in 3-D LGE-MRI},
  author={Kurzendorfer, Tanja and Forman, Christoph and Schmidt, Michaela and Tillmanns, Christoph and Maier, Andreas and Brost, Alexander},
  journal={Computerized Medical Imaging and Graphics},
  volume={59},
  pages={13--27},
  year={2017},
  publisher={Elsevier}
}

@inproceedings{3260-02,
  title={A kernel method for the two-sample-problem},
  author={Gretton, Arthur and Borgwardt, Karsten and Rasch, Malte and Sch{\"o}lkopf, Bernhard and Smola, Alex J},
  booktitle={Advances in neural information processing systems},
  pages={513--520},
  year={2007}
}

@INPROCEEDINGS{3260-03,
    author={T. {Vu} and H. {Jain} and M. {Bucher} and M. {Cord} and P. {Pérez}},
    booktitle={CVPR},
    title={ADVENT: Adversarial Entropy Minimization for Domain Adaptation in Semantic Segmentation},
    year={2019},
    volume={},
    number={},
    pages={2512-2521},
    doi={10.1109/CVPR.2019.00262},
    ISSN={1063-6919},
    month={June},
}

@INPROCEEDINGS{3260-04,
  author={J. {Zhu} and T. {Park} and P. {Isola} and A. A. {Efros}},
  booktitle={ICCV}, 
  title={Unpaired Image-to-Image Translation Using Cycle-Consistent Adversarial Networks}, 
  year={2017},
  volume={},
  number={},
  pages={2242-2251},}

@inproceedings{3260-05,
     author = {Luo, Yawei and Liu, Ping and Guan, Tao and Yu, Junqing and Yang, Yi},
     booktitle = {Advances in Neural Information Processing Systems},
     pages = {20612--20623},
     publisher = {Curran Associates, Inc.},
     title = {Adversarial Style Mining for One-Shot Unsupervised Domain Adaptation},
     volume = {33},
     year = {2020}
}

@ARTICLE{3260-06, 
    author={X. {Zhuang}}, 
    journal={IEEE Trans Pattern Anal Mach Intell}, 
    title={Multivariate mixture model for myocardial segmentation combining multi-source images}, 
    year={2019}, 
    volume={}, 
    number={}, 
    pages={1-1}, 
    ISSN={0162-8828}, 
}

@INPROCEEDINGS{3260-07,
  author={X. {Huang} and S. {Belongie}},
  booktitle={ICCV}, 
  title={Arbitrary Style Transfer in Real-Time with Adaptive Instance Normalization}, 
  year={2017},
  volume={},
  number={},
  pages={1510-1519},}

@INPROCEEDINGS{3260-08,
  author={Y. {Tsai} and W. {Hung} and S. {Schulter} and K. {Sohn} and M. {Yang} and M. {Chandraker}},
  booktitle={CVPR}, 
  title={Learning to Adapt Structured Output Space for Semantic Segmentation}, 
  year={2018},
  volume={},
  number={},
  pages={7472-7481},
}

@InProceedings{3260-09,
author="Vesal, Sulaiman
and Ravikumar, Nishant
and Maier, Andreas",
title="Automated Multi-sequence Cardiac MRI Segmentation Using Supervised Domain Adaptation",
booktitle="STACOM",
year="2020",
pages="300--308",
isbn="978-3-030-39074-7"
}

\end{document}